\documentclass[a4paper]{jpconf}
\usepackage{graphicx}

\begin{document}

\title{Quasi-normal modes of spin-3/2 fields in $D$-dimensional Reissner-Nordstr\"om black hole spacetimes using the continued fraction method}

\author{A. S. Cornell$^1$ and G. E. Harmsen$^2$}
\address{National Institute for Theoretical Physics; School of Physics, University of the Witwatersrand, Johannesburg, Wits 2050, South Africa.}

\ead{$^1$alan.cornell@wits.ac.za, $^2$gerhard.harmsen5@gmail.com}

\begin{abstract}
In a recent paper we calculated the field equations of spin-3/2 fields in a $D$-dimensional Reissner-Nordstr\"om black hole spacetime whilst maintaining the gauge symmetry of the Rarita-Schwinger equation. We were also able to determine the quasi-normal modes of the associated gauge invariant variables using the WKB approximation and the asymptotic iteration method (AIM). However, it was found that for higher dimension, and especially for the near extremal cases, the effective potential developed another maximum. The shape of the potential posed difficulties for the WKB approximations, as well as the AIM. As such, in this proceedings we would like to explore the connection between the AIM and the continued fraction method, and determine a possible reason for the difficulty in calculating the quasi-normal modes for spin-3/2 fields in this spacetime.
\end{abstract}


\section{Introduction}

\par The Rarita-Schwinger equation describes the equations of motion of spin-3/2 fields on a Reissner-Nordstr\"om (RN) spacetime by making use of a supercovariant derivative, that is:
\begin{equation}
\gamma^{\mu\nu\alpha} \tilde{\cal D}_\nu \psi_\alpha = 0 , 
\end{equation}
where 
\begin{equation}
\gamma^{\mu\nu\alpha} = \gamma^\mu \gamma^\nu \gamma^\alpha - \gamma^\mu g^{\nu \alpha} + \gamma^\nu g^{\mu \alpha} - \gamma^\alpha g^{\mu\nu} 
\end{equation} 
is the antisymmetric product of Dirac matrices, $\psi_\alpha$ is the spin-3/2 field and 
\begin{equation}
\tilde{\cal D}_\mu = \nabla_\mu + \frac{1}{2} \sqrt{\frac{D-3}{2(D-2)}} \gamma_\rho F_\mu^\rho + \frac{i}{4\sqrt{2(D-2)(D-3)}}\gamma_{\mu\rho\sigma} F^{\rho \sigma}
\end{equation}
is the supercovariant derivative calculated in Ref.\cite{Chen:2017kqa}. This equation preserves the gauge invariance of the Rarita-Schwinger equation that was noted in Refs.\cite{Chen:2016qii, Chen:2015jga}. However, when attempting to calculate the quasi-normal modes (QNMs) several parameter choices do not lead to convergent or reliable numbers when using the WKB approximation and the asymptotic iteration method (AIM) \cite{Chen:2017kqa}.

\par For the WKB approximation it was found that unreliable QNM results were those where the parameter choices led to higher order terms dominating over the lower order terms. Recall that the WKB approximation is generated from a series expansion, and as such results where higher order terms dominate over lower order ones cannot be considered reliable. As for the AIM, the results did not converge for such choices of parameters, making them unreliable also, where it was found that as the number of iterations in this method were increased, these results did not converge to any one number. As such, in our previous work \cite{Chen:2017kqa}, no numbers for these QNMs were given for such parameter choices.

\par It was proposed that this peculiar behaviour was due to a second maxima developing in the effective potential, where this was clearly visible in higher dimensions ($D > 7$) and when the charge of the black hole approached its extremal value. Given that the WKB approximation, when applied to black hole studies, involves expanding the potential around its maximum value and then matching two WKB solutions from either side of this maximal region of the potential, having two maxima renders the approximation unreliable in its current form. However, it remains unclear as to why the AIM would break down, given all previous investigations have found it to be quite stable \cite{Cho:2011sf, Cho:2009cj, Cho:2009wf}.

\par As such, it is the aim of this proceedings to investigate these unreliable quasi-normal frequencies with a third method for calculating QNMs, that of the continued fraction method (CFM) \cite{Leaver:1985ax}, and to test the claim that the AIM and the CFM are connected \cite{Matamala:2007}.


\section{The spin-3/2 fields equations in a RN spacetime}

\par Using the line element for the RN spacetime,
\begin{equation}
ds^2 = - f dt^2 + \frac{1}{f} dr^2 + r^2 d\bar{\Omega}_{D-2}^2 , 
\end{equation}
where $f = 1 - 2M/r^{D-3} + Q^2/r^{2D - 6}$, and $d\bar{\Omega}_{D-2}$ denotes the metric of the $D-2$ sphere (over-bars will represent terms from this metric), our wave functions for the spin-3/2 fields can be constructed from ``non TT eigenmodes" and ``TT eigenmodes" \cite{Chen:2017kqa, Chen:2016qii}. 

\par Focussing on the non-TT eigenfunctions, where a large number of the unreliable QNMs were observed in Ref.\cite{Chen:2017kqa}, the radial and temporal wave functions can be written as:
\begin{equation}
\psi_r = \phi_r \otimes \bar{\psi}_{(\lambda)} \;\;\;\; \mathrm{and} \;\;\;\; \psi_t = \phi_t \otimes \bar{\psi}_{(\lambda)} , 
\end{equation}
where $\bar{\psi}_{(\lambda)}$ is an eigenspinor on the $D-2$ sphere with eigenvalue $i \bar{\lambda}$ given by $\bar{\lambda} = (j + (D-3)/2)$, where $j = 3/2, 5/2, 7/2, \ldots$. Our angular wave function is written as:
\begin{equation}
\psi_{\theta_i} = \phi_\theta^{(1)} \otimes \bar{\nabla}_{\theta_i} \bar{\psi}_{(\lambda)} + \phi^{(2)}_\theta \otimes \bar{\gamma}_{\theta_i} \bar{\psi}_{(\lambda)} ,
\end{equation}
where $\phi_\theta^{(1)}$, $\phi_\theta^{(2)}$ are functions of $r$ and $t$ which behave like 2-spinors. 

\par Using the Weyl gauge, $\phi_t = 0$, we introduce a gauge invariant variable
\begin{equation}
\Phi = - \left( \frac{\sqrt{f}}{2} i \sigma^3 + \frac{i Q}{2 r^{D-3}} \right) \phi_\theta^{(1)} + \phi_\theta^{(2)},
\end{equation}
to determine the independent equations of motion. Component-wise we can write $\Phi$ as:
\begin{equation}
\Phi = \left( \begin{array}{c} \phi_1 e^{-i\omega t} \\ \phi_2 e^{-i \omega t} \end{array} \right) , 
\end{equation}
where $\phi_1$ and $\phi_2$ are purely radially dependent terms. To simplify the equations we further set
\begin{equation}
\phi_{1}=\frac{\displaystyle \left(\frac{D-2}{2}\right)^{2}f-\left(\bar{\lambda}+C \right)^{2}}{Br^{(D-4)/2}f^{1/4}}\tilde{\phi}_{1}\;\; \;\;\mathrm{and} \;\;\;\; \phi_{2}=\frac{\displaystyle \left(\frac{D-2}{2}\right)^{2}f-\left(\bar{\lambda}-C \right)^{2}}{Ar^{(D-4)/2}f^{1/4}}\tilde{\phi}_{2}.
\end{equation}

\par This leads to a set of decoupled second-order differential equations in the tortoise coordinates, defined by $dr_* = dr/f(r)$:
\begin{eqnarray}
-\frac{d^{2}}{dr_{*}^{2}}\tilde{\phi}_{1} + V_{1}\tilde{\phi}_{1} & = & \omega^{2}\tilde{\phi}_{1}, \\
-\frac{d^{2}}{dr_{*}^{2}}\tilde{\phi}_{2} + V_{2}\tilde{\phi}_{2} & = & \omega^{2}\tilde{\phi}_{2}, \nonumber
\end{eqnarray}
where $V_{1,2} = \pm f(r) dW/dr + W^{2}$ and
\begin{eqnarray}
W & = & \frac{(D-3)\sqrt{f}}{rAB}\Bigg[( \bar{\lambda} + C)\frac{2}{D-2}AB + \frac{D-2}{2}\left(C + \bar{\lambda}\left(1 - f\right)\right) \Bigg] \nonumber \\
&& \hspace{1cm} - \frac{D-4}{r(D-2)}\sqrt{f}(\bar{\lambda}+ C), 
\end{eqnarray}
\begin{equation}
A = \frac{D-2}{2}\sqrt{f} + \left( \bar{\lambda} + C\right), \:\:\:B = \frac{D-2}{2}\sqrt{f} -\left( \bar{\lambda} + C \right)\;\; \mathrm{and} \;\; C = \left(D-2 \right)\frac{Q}{2r^{D-3}}.
\end{equation}


\section{The asymptotic iteration method}

\par In our previous works we had used the AIM to generate some of our QNMs. In the AIM we first single out the asymptotic behaviour, which is due to the QNM boundary condition that the wave function must have the form of $\tilde{\phi}_{1}\sim e^{\pm i \omega r_{*}}$. Our first step is to determine the tortoise coordinates for the specific spacetime we are considering, that is the spacetime for specific values of $Q$ and $D$. Plugging this into the wave function we then transform our wave function to the coordinates 
\begin{equation}
\xi^{2} = 1 - \frac{r_{+}}{r},
\end{equation}
where $r_{+}=\left(M+\sqrt{M^{2}-Q^{2}} \right)^{1/(D-3)}$, such that $\xi \in [0,1]$. Once our wave function has been written in terms of this new coordinate $\xi$ we separate out the asymptotic behaviour and write our wave function as $\tilde{\phi_{1}}=\beta(\xi)\chi(\xi)$, where $\beta(\xi)$ contains the asymptotic behaviour. In the case of $Q=0$ and $D=4$ we have that the wave function is written as
\begin{equation}\label{eq:13}
\tilde{\phi}_{1} = \xi^{4 i M \omega }\left(1-\xi^{2} \right)^{- 2 i M \omega}e^{\frac{2 i M \omega}{1 - \xi^{2}}}\chi(\xi).
\end{equation}
The lowest order coefficients on the AIM can be obtained as:
\begin{eqnarray}
\lambda_{0}& = & 2\frac{\beta'(\xi)}{\beta(\xi)} + \frac{\left(-\left(\xi\right)^{-2}(1-\xi^2)^{2} - 4 (1-\xi^2) \right)f\left(\frac{2M}{1-\xi^{2}}\right) + 4Mf'\left(\frac{2M}{1-\xi^{2}}\right)}{\frac{1}{\xi}(1-\xi^2)^{2}f(\frac{2M}{1-\xi^{2}})} , \nonumber \\
s_{0} & = & \frac{\beta''(\xi)}{\beta(\xi)}+ \frac{\beta'(\xi)}{\beta(\xi)}\left( \frac{\left(-\left(\xi\right)^{-2}(1-\xi^2)^{2} - 4 (1-\xi^2) \right)f\left(\frac{2M}{1-\xi^{2}}\right) + 4Mf'\left(\frac{2M}{1-\xi^{2}}\right)}{\frac{1}{\xi}(1-\xi^2)^{2}f(\frac{2M}{1-\xi^{2}})}\right)\\
&& + \frac{16 M^{2}\xi^2}{f\left(\frac{2M}{1-\xi^{2}}\right)^{2}(1-\xi^2)^{4}}\left(\omega^{2} - V(\frac{2M}{1-\xi^{2}})\right) , \nonumber
\end{eqnarray}
where we have let
\begin{equation}
\beta(\xi) = \xi^{4 i M \omega}\left(1-\xi^{2} \right)^{- 2 i M \omega}e^{\frac{2 i M \omega}{1 - \xi^{2}}},
\end{equation}
and $f(x) = 1 - 2 M/x^{D-3} + Q^2/x^{2D-6}$. As such the AIM differential equation is:
\begin{equation}
\frac{d^2 \chi}{d\xi^2} = \lambda_0 (\xi) \frac{d\chi}{d\xi} + s_0 (\xi) \chi . \label{AIMde}
\end{equation}

\par The higher order $\lambda$ and $s$ can be calculated from the relations:
\begin{equation}
\lambda_{n}=\lambda_{n-1}'+s_{n-1}+\lambda_{0}\lambda_{n-1} \;\; \;\; \mathrm{and} \;\; \;\; s_{n}=s_{n-1}'+s_{0}\lambda_{n-1},
\end{equation}
with the quasi-normal frequencies, $\omega$, being obtained from the equation:
\begin{equation}
s_{n}\lambda_{n+1}-s_{n+1}\lambda_{n}=0.
\end{equation}
Iterating this method for a sufficiently large number of iterations, the QNMs usually become stable, indicating that we have found the $\omega$ we are looking for. However, this was not the case for several modes in our previous work \cite{Chen:2017kqa}. 

\par As was noted in Ref.\cite{Matamala:2007}, the AIM and CFM are closely connected. However, the correspondence espoused by Ref.\cite{Matamala:2007} is overtly simplistic given it only links two of the three terms of the typical CFM recurrence relation to the $\lambda_0$ and $s_0$ of the AIM. The third term remains unrelated (see Eq. (4) of Ref.\cite{Leaver:1985ax}), despite its key role (see $\alpha_n$ defined below). Furthermore, it was noted in the earlier works on the CFM (see Ref.\cite{Leaver:1985ax, Gautschi}), that the coefficients should only depend on parameters, not variables, as is the case for the $\lambda_0$ and $s_0$ defined in this scenario.


\section{The continued fraction method}

\par To analyse the connection between the CFM and AIM, we first recall that in the original work on the CFM, where a power series expansion of the wave function was done after the removal of asymptotic behaviour (cf. Eq. (\ref{eq:13}) above), that is:
\begin{equation}
\chi (\xi) = \sum_{n=0}^\infty a_n \left( 1 - \frac{1}{\xi} \right)^n ,
\end{equation}
where $\chi$ is the wave function of the AIM differential equation, Eq. (\ref{AIMde}). The expansion coefficients, $a_n$, are then defined by a three-term recurrence relation:
\begin{equation}
\alpha_n a_{n+1} + \beta_n a_n + \gamma_n a_{n-1} = 0 , \;\; n = 1, 2, \ldots \label{LeaverEq4}
\end{equation}
The recurrence coefficients are in this case:
\begin{eqnarray}
\alpha_n & = & n(n+1) , \nonumber\\
\beta_n & = & - n \xi (2 + \lambda_0 \xi ) , \label{recurrence-coefficients} \\
\gamma_n & = & - s_0 \xi^4 \nonumber
\end{eqnarray}
Note though that these coefficients should not depend on $\xi$ \cite{Gautschi}, and furthermore $\gamma_n$ has no $n$ dependence, yet in Ref.\cite{Matamala:2007} it was the $\alpha_n$ which was undefined in their correspondence (though we do see here that it is unrelated to the $\lambda_0$ and $s_0$ functions of the AIM). Furthermore, due to the complicated functionality of $\lambda_0$ and $s_0$, it would seem impossible to remove all $\xi$ dependencies from these coefficients, in general, even for other possible power series expansions. This could be why the AIM could not give reliable QNMs for higher dimension and near extremal cases, as these represent the most complicated forms for $\lambda_0$ and $s_0$.

\par If it were possible to computer the quasi-normal frequencies, they would be the complex values of $\omega$ for which the series defined by \cite{Leaver:1985ax}
\begin{equation}
F(\omega) = - \frac{\gamma_1}{\beta_1 - \displaystyle \frac{\alpha_1 \gamma_2}{\beta_2 - \displaystyle \frac{\alpha_2 \gamma_3}{\beta_3 - \ldots}}} ,
\end{equation}
converges uniformly as $\xi \to \infty$ (where the convergence of this series is a separate issue from the convergence of the continued fraction). This analytic function is empirically found to converge for all $\omega$ that are not purely positive imaginary. This restriction is related to the absence of a minimal solution to recurrence relations, Eq. (\ref{LeaverEq4}) \cite{Leaver:1985ax}. 

\par Now when $\omega$ is a QNM, $\omega_n$, the sequence of the expansion coefficients is the minimal solution to the recurrence relation, Eq. (\ref{LeaverEq4}), and the ratio of the first two expansion coefficients is equal to the value of this continued fraction:
\begin{equation}
\frac{a_1 (\omega_n )}{a_0 (\omega_n)} = F(\omega_n) . 
\end{equation}
Since this ratio is also given for any $\omega$, we then have the equation
\begin{equation}
F(\omega_n ) = - \frac{\beta_0 (\omega_n)}{\alpha_0 (\omega_n )} , 
\end{equation}
which holds whenever $\omega$ is a QNM. However, whenever $\omega$ is not a QNM (and also not purely positive imaginary), the continued fraction still converges and the expression 
\begin{equation}
\frac{\beta_0 (\omega)}{\alpha_0(\omega)} + F(\omega)
\end{equation}
is an analytic function of $\omega$ whose zeroes are the QNMs. Note that being analytic makes this expression an ideal target for a numerical root search.

\par However, that the recurrence coefficients cannot seemingly be found in a $\xi$ independent way means the whole CFM is not applicable \cite{Gautschi}. And whilst we have shown that the AIM and CFM can be related, though not as trivially as espoused in Ref. \cite{Matamala:2007}, this correspondence breaks down when the recurrence coefficients do not depend solely on the parameters, but retain a variable dependence. This breakdown of a correspondence between the AIM and the CFM could explain the inability of the AIM to converge in this case, and warrants further investigation.

\section*{Acknowledgements}

\noindent ASC and GEH are supported in part by the National Research Foundation of South Africa. We also wish to acknowledge the useful discussions with Prof. Hing-Tong Cho and Dr. Chun-Hung Chen, during the production of this short note.


\end{document}